\PassOptionsToPackage{table}{xcolor}
\documentclass{article}
\usepackage{spconf,amsmath,graphicx,xcolor}
\usepackage{hyperref} 
\usepackage{booktabs}
\usepackage{multirow}
\usepackage{ifthen}
\usepackage{tabularx}
\usepackage{soul}
\usepackage{tablefootnote}
\usepackage{amssymb,xcolor,stackengine,graphicx}

\definecolor{bleudefrance}{rgb}{0.19, 0.55, 0.91}
\newcolumntype{Y}{>{\centering\arraybackslash}X}
\setul{0.5ex}{0.3ex}

\title{AV-SUPERB: A Multi-task Evaluation Benchmark \\for Audio-visual Representation Models }

\name{\begin{tabular}[c]{@{}c@{}c@{}c@{}c@{}c@{}}
    Yuan Tseng$^{1}$, 
    Layne Berry$^{2*}$, 
    Yi-Ting Chen$^{3*}$, 
    I-Hsiang Chiu$^{1*}$,
    Hsuan-Hao Lin$^{1*}$, 
    Max Liu$^{1*}$, \\ 
    Puyuan Peng$^{2*}$, 
    Yi-Jen Shih$^{1*}$,
    Hung-Yu Wang$^{1*}$,
    Haibin Wu$^{1*}$,
    Po-Yao Huang$^{4}$, 
    Chun-Mao Lai$^{1}$, \\ 
    Shang-Wen Li$^{4}$, 
    David Harwath$^{2}$, 
    Yu Tsao$^{3}$,
    Abdelrahman Mohamed$^{5}$,
    Chi-Luen Feng$^{1}$, 
    Hung-yi Lee$^{1}$ 
    \thanks{$^*$ Equal contribution; sorted alphabetically.}
    \end{tabular}
}

\address{
    $^{1}$ National Taiwan University, Taiwan %
    $^{2}$ University of Texas at Austin, USA \\
    $^{3}$ Academia Sinica, Taiwan %
    $^{4}$ Meta AI
    $^{5}$ Rembrand \\ 
    \texttt{r11942082@ntu.edu.tw} 
}

\begin{document}
\ninept
\maketitle
\begin{abstract}
Audio-visual representation learning aims to develop systems with human-like perception by utilizing correlation between auditory and visual information. 
However, current models often focus on a limited set of tasks, and generalization abilities of learned representations are unclear.
To this end, we propose the AV-SUPERB benchmark that enables general-purpose evaluation of unimodal audio/visual and bimodal fusion representations on 7 datasets covering 5 audio-visual tasks in speech and audio processing.
We evaluate 5 recent self-supervised models and show that none of these models generalize to all tasks, emphasizing the need for future study on improving universal model performance.
In addition, we show that representations may be improved with intermediate-task fine-tuning and audio event classification with AudioSet serves as a strong intermediate task.
We release our benchmark with evaluation code\footnote{\href{https://github.com/roger-tseng/av-superb}{https://github.com/roger-tseng/av-superb}} and a model submission platform\footnote{\href{https://av.superbbenchmark.org}{https://av.superbbenchmark.org}}\addtocounter{footnote}{-1}\addtocounter{Hfootnote}{-1} to encourage further research in audio-visual learning.
\end{abstract}
\section{Introduction}
\label{sec:intro}

Emulating the seamless integration of multiple tasks in human cognition, such as spoken language comprehension, sound event detection, and visual object recognition has been a long-standing goal of computational research.
Prior research demonstrates that the pretrain-then-finetune paradigm is an effective and scalable method of building multitasking algorithmic systems for speech \cite{wav2vec2,hubert}, audio \cite{byola,amae}, and vision \cite{vit,mae}. 
In the pretraining stage, models can often learn meaningful representations from unlabelled data alone through optimization of contrastive, masked prediction, or other self-supervised loss functions. 
These pretrained representations can then be applied to diverse tasks just by fine-tuning minimal additional parameters.

In order to better measure progress in representation learning, previous works have established multitask benchmarks in speech \cite{superb,superbsg}, audio \cite{hear}, and vision \cite{sslvideo1,sslvideo2}.
However, these benchmark predominantly evaluate performance in isolation within single modalities.
This approach overlooks the inherent multimodal nature of human perception, which synergistically integrates auditory and visual cues \cite{mcgurk,GHAZANFAR2006278}.
While audio-visual representation learning has made significant progress \cite{soundnet,kinetics,avts,xdc,patrick21}, 
the assessment of these models tends to be task-specific, leaving the broader generalization capabilities across various audio-visual challenges less understood.
This complicates comparitive analysis of different models and training strategies, impeding the development of more robust and versatile audio-visual representation learning approaches.

To address this issue, we propose AV-SUPERB, a standardized benchmark for comprehensively evaluating representations across seven distinct datasets involving five speech and audio processing tasks.
AV-SUPERB comprises of three tracks to assess audio, video, and audio-visual fusion representations. 
We envision that these distinct tracks will allow researchers in speech, audio, and video representation learning alike to compare learning strategies across models and modalities, enabling broader analysis of their effectiveness.

Our contributions are four-fold: 
(1) \textbf{Diverse-domain evaluation}: 
We propose the first audio-visual learning benchmark that encompasses multiple datasets and tasks, covering both speech and audio domains.
(2) \textbf{Easy and reproducible benchmarking}: 
We release evaluation code and a dedicated model submission platform 
that ensures reproducible evaluation on dynamic Youtube datasets and reduces computational entry barriers.
(3) \textbf{Intermediate-task fine-tuning}:
Our work emphasizes the potential benefits of full fine-tuning on intermediate tasks for improving performance on out-of-domain downstream tasks. 
(4) \textbf{Layer-wise analysis}: 
We show that different layers contribute variably to task performance, suggesting that simply using representations of the final layer is suboptimal, motivating the weighted-sum evaluation approach.

\begin{figure}[tb]
  \centering
  \includegraphics[width=1.0\linewidth]{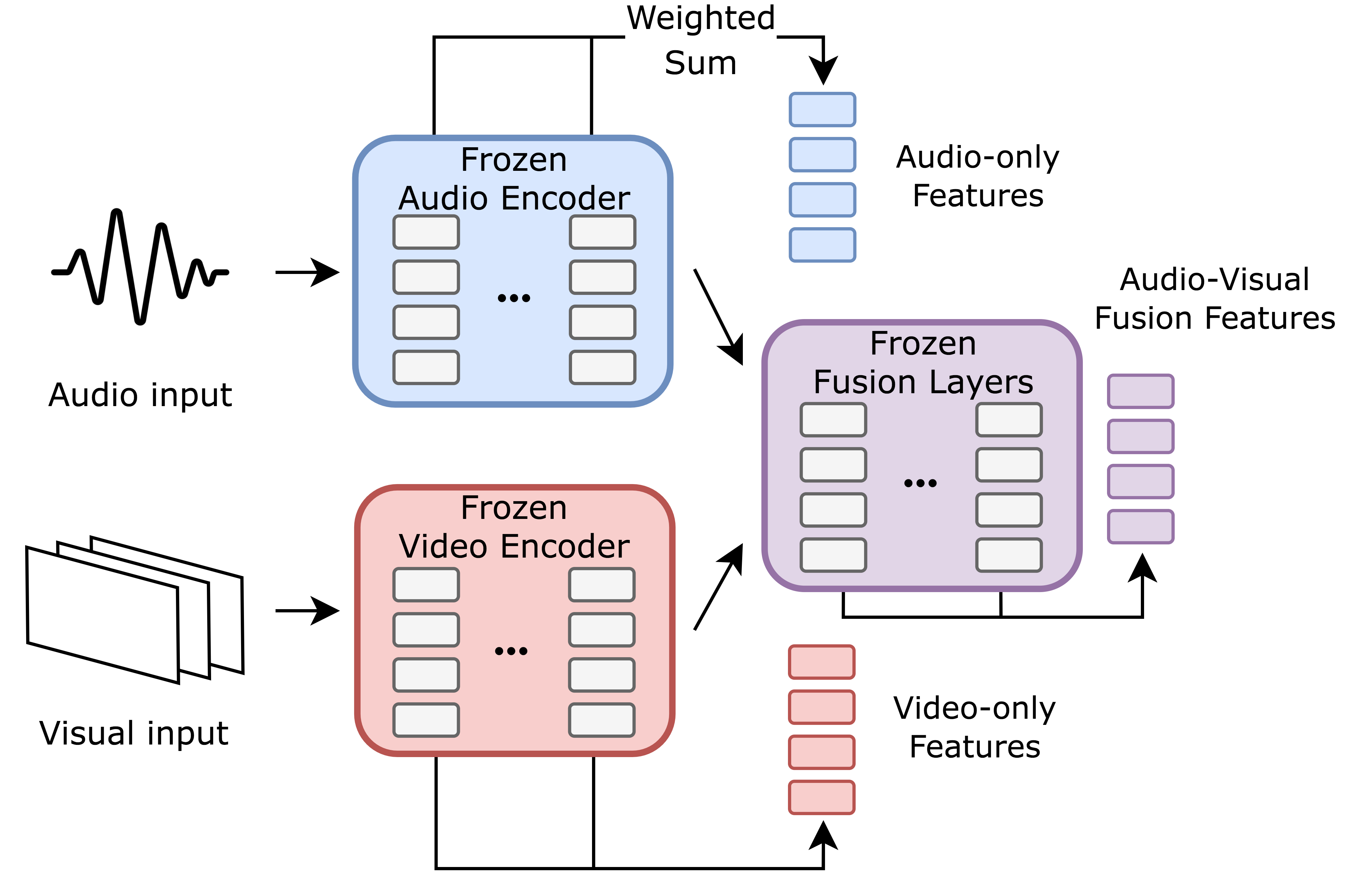}
  \vspace{-7mm}
  \caption{
  We consider three evaluation scenarios: extracting features using inputs from one or both modalities.
  Following \cite{superb}, the weighted-sum of features from Transformer layers (if applicable) are used as input for fine-tuning a small downstream model for each individual task.
  Details of selected tasks are given in Section ~\ref{section:tasks}. 
  }
  \vspace{-3mm}
  \label{fig:framework}
\end{figure}

\vspace{-2mm}
\section{Related Work}
\label{sec:related}

Recognizing how the close relation between audition and vision facilitates multimodal human perception, many audio-visual datasets have been gathered for action recognition \cite{kinetics400,kinetics,ucf101,epickitchen}, speech recognition \cite{grid,lrs,lrs3}, speaker recognition \cite{voxceleb,voxceleb2}, and a variety of other tasks to study audio-visual learning.
However, most models are trained and evaluated on different datasets with different experiment settings, which increases comparison difficulty and obfuscates the broad applicability of proposed methods.
Hence in the AV-SUPERB benchmark, we select a diverse set of datasets from multiple tasks to comprehensively compare works in audio-visual representation learning.

Past multitask benchmarks in speech \cite{superb,superbsg}, audio \cite{hear}, and video representation learning \cite{feichtenhofer,sslvideo1,sslvideo2} allow for fairer comparison of different models and promote research towards general approaches that are applicable to a variety of real-world tasks.
SUPERB \cite{superb,superbslt} and SUPERB-SG \cite{superbsg} evaluate speech representation models on a wide range of downstream tasks covering content, speaker, and other different aspects of speech.
Additionally, the HEAR benchmark \cite{hear} evaluates audio representations in diverse domains beyond speech, such as music and environmental sounds.
For video representations, the SEVERE-benchmark \cite{sslvideo1} compares video self-supervised learning models on a diverse set of datasets to measure model sensitivity to different properties of downstream tasks. %
Feichtenhofer et al. \cite{feichtenhofer} extend 4 image self-supervised learning methods to video representations and compare their efficacy on several downstream datasets,
while Kumar et al. \cite{sslvideo2} focus on the effects of different factors in self-supervised video pretraining.
However, these works focus on individual domains and cannot make use of the relationship between paired audio/visual inputs. 

Previous multitask multimodal benchmarks focus on egocentric videos \cite{ego4d}, vision-and-language domains \cite{vlue,value} or general multimodal learning\cite{multibench}.
In contrast, AV-SUPERB specializes in audio-visual tasks from speech and audio processing, allowing for more holistic assessment of representation models of audio and video alike.

\vspace{-3mm}
\section{Benchmark Details}
\label{sec:details}

\begin{table*}[htb]
\renewcommand{\arraystretch}{1.1}
\centering
\setlength\tabcolsep{4pt} 
\begin{tabular}{ccc cccc ccc}
\toprule
\multirow{4}{*}[-10pt]{Representation Type} & \multirow{4}{*}[-10pt]{Params.} & \multirow{4}{*}[-10pt]{\begin{tabular}{c} Overall \\ Score\end{tabular} } & \multicolumn{4}{c}{Audio-Visual} & \multicolumn{3}{c}{Speech-Visual}\\
\cmidrule(lr){4-7}
\cmidrule(lr){8-10}
& & & \multicolumn{2}{c}{AEC} & \multicolumn{2}{c}{AR} & ASR & ASV & ER\\ 
\cmidrule(lr){4-5}
\cmidrule(lr){6-7}
\cmidrule(lr){8-8}
\cmidrule(lr){9-9}
\cmidrule(lr){10-10}
& & & AS-20K & VGGSound & \begin{tabular}{c} Kinetics-\\Sounds\end{tabular} & UCF101 & LRS3-TED & VoxCeleb2 & IEMOCAP \\
& & & (mAP $\uparrow$) & (Acc. $\uparrow$) & (Acc. $\uparrow$) & (Acc. $\uparrow$) & (CER $\downarrow$) & (EER $\downarrow$) & (Acc. $\uparrow$) \\
\midrule
\multicolumn{1}{l}{\textit{Audio-only}} & & & & & & & \vspace{2pt}\\
FBANK & 0 & 36.88
& 2.8	& 7.76	& 24.73	& 19.91	& 21.43 & 27.16	& 51.52 \\
HuBERT & 95M & 53.66
& 14.3          & 30.21         & 51.46         & 36.06         & \textbf{2.96} & \underline{15.58}             & \textbf{62.14} \\
AV-HuBERT* & 90M & 53.20
& 12.6          & 31.14         & 49.02         & 38.58         & \underline{3.01}          & \textbf{14.45}    & 58.54  \\
RepLAI & 5M & 39.70
& 12.3          & 27.01         & 45.90         & 33.85         & 66.09         & 32.58             & 57.53   \\
AVBERT & 10M & 44.81
& \underline{20.5}	& \underline{37.67}	& \underline{55.28}	& \underline{43.26}	& 80.23 & 23.74	& \underline{60.94} \\
MAViL & 86M & 54.11
& \textbf{21.6} & \textbf{39.91} & \textbf{57.28} & \textbf{45.68} & 24.43         & 20.71             & 59.46   \\
\midrule
\multicolumn{1}{l}{\textit{Video-only}} & & & & & & & \vspace{2pt}\\
HoG & 0 & 25.39
& 1.5	& 3.81	& 18.70	& 25.67 & 71.46	& 36.32		& 35.83 \\
AV-HuBERT* & 103M & 33.48
& 2.4           & 5.90          & 24.73         & 37.55         & \textbf{50.91}  & \textbf{11.90}  & 26.59  \\
RepLAI & 15M & 36.40
& 5.5          & 13.5          & 46.68         & 56.69          & \underline{71.33}           & 36.95           & 40.72   \\
AVBERT & 37M & 47.69
& \underline{11.5}	& \underline{28.73}	& \underline{62.67}	& \underline{77.42}	& 72.29	& \underline{20.00}	& \textbf{45.8} \\
MAViL & 87M & 49.70
& \textbf{18.0} & \textbf{32.08}  & \textbf{74.01} & \textbf{79.37} & 74.03         & 24.58  & \underline{43.03}   \\
\midrule

\multicolumn{1}{l}{\textit{Audio-visual fusion}} & & & & & & & \vspace{2pt}\\
AV-HuBERT & 103M & 53.42
&  13.3         & 32.69         & 52.23         & 41.46           & \textbf{2.75} & \textbf{9.46}   & 46.45   \\
AVBERT & 43M & 54.85
& \underline{22.9}	& \underline{44.54}	& \underline{71.31}	& \underline{71.76}	& 70.12 & \underline{18.31}		& \textbf{61.87} \\
MAViL & 187M & 62.36
& \textbf{26.7} & \textbf{47.22} & \textbf{79.51} & \textbf{77.98}  & \underline{30.18}         & 19.67           & \underline{54.94}   \\
\bottomrule
\multicolumn{10}{l}{\footnotesize *In order to fairly compare HuBERT \& AV-HuBERT, we set features of the opposing modality to 0 and extract features from the 12-layer fusion } \\
\multicolumn{10}{l}{\footnotesize Transformer for audio-only and video-only tracks.}
\vspace{-3mm}
\end{tabular}
\caption{Main results. Best results for each track are highlighted in bold. Second-best results are underlined.
We observe that MAViL excels at audio processing tasks, while HuBERT and AV-HuBERT are better for speech processing tasks.
}
\label{table:main}
\vspace{-5mm}
\end{table*}

As shown in Figure~\ref{fig:framework}, audio-visual models typically consist of two separate unimodal encoders followed by multimodal fusion layers. 
Based on this design, we setup three evaluation tracks in AV-SUPERB to benchmark representations from the two encoder and fusion layers, referred as audio-only, video-only, and audio-visual fusion features.
This also allows for easy comparison with previous unimodal representation models.

Instead of striving for best possible performance for each task, the goal of our benchmark is to provide insight on the generalization capabilities of pretrained representations; 
therefore, we freeze the parameters of the task-invariant pretrained representation model (hereby referred as upstream model), and only fine-tune the parameters of the task-specific model (hereby referred as downstream model), following previous work \cite{superb}.
Downstream models are designed to be simple and lightweight in order to purely evaluate representation abilities.
Following the spirit of representation evaluation, we also limit hyperparameter tuning for downstream tasks. 
Although, we recognize that different representations may have vastly different loss landscapes, hence we search for the best performing learning rate from $10^{-1}$ to $10^{-5}$ in log-scale.

\vspace{-2mm}
\subsection{Downstream Task Selection}
\label{section:tasks}

To keep computational costs reasonable, we mainly focus on utterance-level classification tasks in speech and audio processing, with the addition of ASR. 

For audio processing, we select two audio classification tasks that highlight the relevance of different modalities, audio event classification ({AEC}) and action recognition ({AR}). 
Since audio events are often directly caused by actions, these tasks are complementary, and utilizing both audio and visual information can lead to better representations. 
This enables the possibility of learning better representations from multimodal input compared to unimodal baselines.

For speech processing, we select three audio-visual speech processing tasks where visual information is known to be beneficial \cite{lipread1,asv2,er}, automatic speech recognition ({ASR}), automatic speaker verification ({ASV}), and emotion recognition ({ER}), in order to assess model capabilities on three fundamental aspects of speech: content, speaker, and paralinguistic information.

In designing the architecture for the downstream models, we generally follow the setup used for utterance-level tasks in the SUPERB benchmark. Specifically, the downstream model consists of a two-layer fully-connected network. 
This network takes the mean of features extracted from the frozen upstream model as input, and outputs class probabilities.
However, as we also include the frame-level ASR task, we employ a two-layer BiLSTM model that takes the whole representation sequence as input and outputs characters.

\vspace{-3mm}
\subsection{Pretrained Upstream Models}

To showcase the utility of our benchmark, we opt for the base version of four audio-visual upstream models, AV-HuBERT \cite{avhubert}, RepLAI \cite{replai}, Lee et al.'s model \cite{avbert} (hereby referred as AVBERT throughout this paper), and MAViL \cite{mavil}. 
These models were specifically chosen because they each excel at different tasks, underscoring the current gap in multi-tasking capabilities within existing audio-visual models.
They vary substantially in terms of architecture, training objectives, and preprocessing techniques.
We also conduct experiments on the base HuBERT \cite{hubert} model, an unimodal speech representation model with similar design as AV-HuBERT, to make a fairer comparison between audio \& audio-visual features.

Additionally, we incorporate two baselines that use handcrafted features as input for downstream models. 
Specifically, we employ log mel filterbank (FBANK) for audio and histogram of oriented gradients (HoG) for video, respectively.

\vspace{-1mm}
\section{Experimental Results and Discussion}

Previous work has shown that simply using representations extracted at the last layer of a frozen self-supervised model often results in suboptimal performance \cite{pasad2021layer,superb}. 
Hence, we take a learnable weighted-sum of representations extracted over different Transformer layers as the final representation for each downstream task.
For the audio-only and video-only tracks, only unimodal input and the relevant layers are used for extracting representations.
For the audio-visual fusion track, both of the unimodal encoders plus fusion layers are used. 
As the size of representations extracted from fusion Transformer layers differ from those of unimodal layers, we take the weighted-sum for fusion Transformer layers only.

\vspace{-2mm}
\subsection{Downstream Datasets and Training Details}

Evaluation results for the three tracks are given in Table ~\ref{table:main}. For AEC, we evaluate on AudioSet \cite{audioset} and VGGSound \cite{vggsound}, and for AR, we select Kinetics-Sounds \cite{kinetics} and UCF101 \cite{ucf101}. 
Notably, in VGGSound and Kinetics-Sounds, audio and visual information are more correlated.
This is reflected in our results, as audio-visual fusion results in larger gains compared to AudioSet and UCF101.
We report testing set mean average precision for multi-label classification on AudioSet, and accuracy for the remaining three datasets.

For speech processing, we choose LRS3-TED for ASR, VoxCeleb2 for ASV, and IEMOCAP for ER.
For ASR, we optimize CTC loss for character-level ASR, and report character error rate.
For ASV, we first train for speaker identification on a subset of the dev split, then calculate cosine similarity to do verification on the test split and report equal error rate.
For ER, we follow the conventional evaluation policy of removing unbalanced classes to perform four-way classification (neutral, happy, sad, angry) and report accuracy.
Additional details related to datasets and training are given on our submission platform\footnotemark.
\vspace{-2mm}
\subsection{Overall Results}

We find that existing models generally obtain large gains over handcrafted features, yet none of the five models tested were able to outperform all others in every task. 
To gauge universal performance across tasks, we provide an overall score calculated as the mean of either task-specific accuracies or the complement of error rates. 

For the three speech processing tasks (ASR, ASV, ER), AV-HuBERT performs the best on ASR and ASV, and HuBERT achieves superior performance on ER.
Notably, the unimodal HuBERT scores competitively on ASR and ASV as well, despite not being trained to utilize any visual grounding information.

For the four audio processing datasets, MAViL and AVBERT consistently outperforms all other models in all three tracks.
We hypothesize that this is largely due to the diversity and large size of AudioSet data used for pretraining. 
Despite the domain mismatch, AVBERT also performs competitively for the ASV and ER speech tasks, especially in the audio-visual fusion track.

However, MAViL and AVBERT cannot perform ASR well, as simply using handcrafted FBANK features achieves lower error rates.
Comparing their scores in the audio-only and fusion tracks, we see that their fusion layers are unable to effectively utilize the additional lip reading information, as performance is reduced when video is provided.

\begin{table*}[htb]
\resizebox{0.999\textwidth}{!}{
\renewcommand{\arraystretch}{1.07}
\centering
\setlength\tabcolsep{3pt} 
\begin{tabularx}{1.05\textwidth}{Yc cccc ccc}
\toprule
\multirow{4}{*}{} & \multirow{4}{*}[-5pt]{ \begin{tabular}{c} Intermediate Task \\ Fine-tuning Data\end{tabular} }  & \multicolumn{4}{c}{Audio-Visual} & \multicolumn{3}{c}{Speech-Visual}\\
\cmidrule(lr){3-6}
\cmidrule(lr){7-9}
& 
& \multicolumn{2}{c}{AEC} & \multicolumn{2}{c}{AR} & ASR & ASV & ER\\ 
\cmidrule(lr){3-4}
\cmidrule(lr){5-6}
\cmidrule(lr){7-7}
\cmidrule(lr){8-8}
\cmidrule(lr){9-9}
& & \setulcolor{green}\ul{AS-20K} & VGGSound & Kinetics-Sounds & UCF101 &  \setulcolor{purple}\ul{LRS3-TED} & VoxCeleb2 & IEMOCAP \\
& & (mAP $\uparrow$) & (Acc. $\uparrow$) & (Acc. $\uparrow$) & (Acc. $\uparrow$) & (CER $\downarrow$) & (EER $\downarrow$) & (Acc. $\uparrow$) \\
\midrule
\multicolumn{1}{l}{\textit{AV-HuBERT}} & & & & & & & & \\
Audio   &   %
& 12.6(\textcolor{red}{-0.6})
& 22.83(\textcolor{red}{-8.31})
& 38.19(\textcolor{red}{-10.83})
& 28.70(\textcolor{red}{-9.88})
& 13.89(\textcolor{red}{-10.88})
& 22.38(\textcolor{red}{-7.93})
& 53.92(\textcolor{red}{-4.62})   \\
Video   &   %
& 2.5(\textcolor{bleudefrance}{+0.1})
& 6.12(\textcolor{bleudefrance}{+0.22})
& 25.35(\textcolor{bleudefrance}{+0.62})
& 42.03(\textcolor{bleudefrance}{+4.48})
& 35.48(\textcolor{bleudefrance}{+15.43})
& 11.40(\textcolor{bleudefrance}{+0.50})
& 32.69(\textcolor{bleudefrance}{+6.10})   \\
Fusion  &  \multirow{-3}{*}{ \setulcolor{purple}\begin{tabular}{c} \ul{LRS3-TED} \\ (video-text pairs)\end{tabular} }
& 5.1(\textcolor{red}{-8.2})
& 17.11(\textcolor{red}{-15.58})
& 38.52(\textcolor{red}{-13.71})
& 40.74(\textcolor{red}{-0.72})
& 22.66(\textcolor{red}{-19.91})
& 11.35(\textcolor{red}{-1.89})
& 43.58(\textcolor{red}{-2.87}) \\
\midrule
\multicolumn{1}{l}{\textit{MAViL}} & & & & & & & & \\
Audio  &  %
& 28.3(\textcolor{bleudefrance}{+6.7}) 
& 44.79(\textcolor{bleudefrance}{+4.89}) 
& 62.93(\textcolor{bleudefrance}{+5.65}) 
& 50.10(\textcolor{bleudefrance}{+4.42}) 
& 23.99(\textcolor{bleudefrance}{+0.44}) 
& 21.77(\textcolor{red}{-1.06}) 
& 58.17(\textcolor{red}{-1.29})   \\
Video   & 
& 20.9(\textcolor{bleudefrance}{+2.9}) 
& 36.68(\textcolor{bleudefrance}{+4.58}) 
& 77.39(\textcolor{bleudefrance}{+3.38}) 
& 86.93(\textcolor{bleudefrance}{+7.56}) 
& 78.59(\textcolor{red}{-4.56})
& 23.93(\textcolor{bleudefrance}{+0.65}) 
& 39.15(\textcolor{red}{-3.88})   \\
Fusion  &  \multirow{-3}{*}{ \setulcolor{green}\ul{AudioSet-2M} }
& 39.1(\textcolor{bleudefrance}{+12.4}) 
& 55.94(\textcolor{bleudefrance}{+8.72}) 
& 84.93(\textcolor{bleudefrance}{+5.42}) 
& 88.07(\textcolor{bleudefrance}{+10.09}) 
& 30.65(\textcolor{red}{-0.47}) 
& 18.61(\textcolor{bleudefrance}{+1.06}) 
& 46.35(\textcolor{red}{-8.59})   \\
\bottomrule
\end{tabularx}
}
\vspace{-2mm}
\caption{Intermediate-task fine-tuning does not generally improve performance across all tasks. Results after intermediate-task fine-tuning (left) and absolute improvements compared to the original self-supervised model (right) are shown. Fine-tuning data for each model is color-coded to the corresponding downstream dataset. }
\label{table:finetune}
\vspace{-3mm}
\end{table*}

\vspace{-2mm}
\subsection{When does Visual Grounding Improve Audio Representation Learning?}

Compared to unimodal audio representation models, audio-visual models may take advantage of information learned from visual grounding to improve audio representations even when only audio input is available at inference. 
Of the five selected models, HuBERT and AV-HuBERT use similar architectures and optimize the same masked cluster prediction objective using k-means clusters of MFCC features as initial targets. 
Although HuBERT is only trained on unimodal speech data, AV-HuBERT is trained to predict \textit{multimodal} cluster targets obtained from both audio and visual modalities. 
By comparing their results on the audio-only track, we see that visual grounding information from multimodal cluster prediction improves representations for VoxCeleb2, VGGSound and UCF101. 

\vspace{-2mm}
\subsection{Layer-wise Contribution Analysis}

After fine-tuning the learnable weighted-sum over all upstream model layers on a downstream task, we may compare layer utilization by examining the weights of each layer in the weighted-sum. \cite{wavlm}
Since the magnitude of representations from each layer may differ, we normalize layer weights for each layer by multiplying the weight with the L2-norm of representation values on the training set.

For MAViL, we find the layers that are commonly more dominant are the last three layers in the audio encoder, and the last two layers in the video encoder and fusion layers.
Despite this, we observe an exception for emotion recognition on IEMOCAP. 
For IEMOCAP, the most dominant layer is the $0^{\text{th}}$ layer instead.

For AV-HuBERT, the final layer often contributes little. 
In the audio-only setup, we see that the layer with the most contribution is the penultimate layer for most speech and audio tasks besides ASR. For ASR, the last two layers are highly dominant on all three tracks.
For non-ASR tasks, we note that when additional visual inputs are given, prior layers increase in contribution only when audio-visual fusion outperforms audio-only performance for AV-HuBERT (VGGSound, Kinetics-Sound, UCF101, VoxCeleb2), suggesting that prior layers in AV-HuBERT are more related to \textit{visual} information, while the last few layers contain more \textit{audio} information.

Overall, the variation in layer usage for different tasks, models, and modalities strongly motivates the use of the learnable weighted-sum technique for evaluation, instead of sub-optimally evaluating the final layer alone.

\vspace{-2mm}
\section{How does intermediate-task fine-tuning affect performance?}
\vspace{-2mm}
\label{sec:inter}

Studies in natural language processing show that pretrained language models can be improved by initial fine-tuning on an intermediate task, followed by further fine-tuning on the target task \cite{stilts,wang-etal-2019-tell}.

In previous sections, we focus on assessing models pretrained in a self-supervised manner. 
However, model creators often release models variants that are fine-tuned further for performing specific downstream tasks. 
For example, MAViL adds 3 Transformer fusion layers after the audio and video encoders, and the whole model is fine-tuned on (audio\&video, class) pairs for audio event classification.
We hypothesize that these supervised models variants may provide improved representations for speech/audio tasks after intermediate-task training.

In order to support our hypothesis, we additionally evaluate fully fine-tuned variants of AV-HuBERT and MAViL on our benchmark, to determine when intermediate-task fine-tuning is beneficial. 
The variant of AV-HuBERT uses the same architecture, and is fine-tuned on 433 hours of (video, text) pairs from LRS3-TED to perform visual speech recognition, whereas the MAViL variant is fine-tuned on the entirety of AudioSet-2M. 
Experiment results are shown in Table ~\ref{table:finetune}.

For AV-HuBERT, we see that visual speech recognition on LRS3-TED is not a suitable intermediate task in general. 
Video-only representations obtain small gains in generalizability, at the cost of greatly reducing audio-only and fusion performance.
We posit that intermediate-task fine-tuning with (video,text) pairs shifts AV-HuBERT Transformer layers to favor video input alone, reducing usability for audio-only and audio-visual inputs.

Contrarily, for audio-visual fusion with MAViL, we see that intermediate-task training on AudioSet-2M not only brings substantial improvements to all AEC and AR datasets, but also improves ASV while maintaining ASR performance.
This suggests that fine-tuning on AudioSet-2M may be sufficiently diverse to improve speaker separability of representations without much loss of content information.

\vspace{-5mm}
\section{Conclusions}
\vspace{-2mm}
\label{sec:conclusion}

We introduce AV-SUPERB, a benchmark for assessing general-purpose capabilities of audio-visual representations. 
AV-SUPERB includes a suite of 7 speech and audio processing datasets covering 5 audio-visual tasks.
The benchmark is split into three tracks: two unimodal audio-only or video-only representations tracks, as well as a bimodal audio-visual fusion track.
This enables easy comparison between unimodal and bimodal learning.
Despite advances made in recent years, our experiments show that none of the models tested generalize to all tasks, leading us to conclude that further research is required to develop universal audio-visual models.

As discussed in Section ~\ref{section:tasks}, although our benchmark aims to comprehensively evaluate audio-visual models, only a limited set of tasks and datasets are included in its current form.
For future work, we wish to incorporate more tasks relevant to additional facets of audio-visual processing, such as cross-modal retrieval, audio-visual localization, and sound/video generation, as well as improving the diversity and comprehensiveness of data sources.

\noindent\textbf{Acknowledgement}: 
We would like to thank Shinji Watanabe for his valuable comments.
We also thank the National Center for High-performance Computing (NCHC) of National Applied Research Laboratories (NARLabs) in Taiwan for providing computational and storage resources.

\vfill\pagebreak
\bibliographystyle{IEEEbib}
\bibliography{strings,refs}

\end{document}